\documentclass{vgtc}                          

\graphicspath{{figures/}{pictures/}{images/}{./}} 

\usepackage{times}                     

\usepackage{subcaption}
\usepackage{pifont}
\usepackage{booktabs}
\usepackage{multirow}
\usepackage{colortbl}
\usepackage{array}
\usepackage{siunitx}
\usepackage{tabulary}
\usepackage{float}

\onlineid{0}

\vgtccategory{Research}

\vgtcinsertpkg

\title{\vid{}: Rapid Assisted Reality Prototyping using Design-Blended Videos}

\author{
Ashwin Ram\thanks{e-mail:ashwinram@u.nus.edu}\\ %
\scriptsize National University of Singapore %
\and Yue Gu\thanks{e-mail:yuegu0825@gmail.com }\\ %
\scriptsize Tsinghua University %
\and Bowen Wang\thanks{e-mail:saltyp0rridge20@gmail.com }\\
\scriptsize Tsinghua University
\and Sneha Jaikumar\thanks{e-mail:sneha.jaikumar@gmail.com }\\ %
\scriptsize National University of Singapore
\and Youqi Wu\thanks{e-mail:wuyoki1203@gmail.com }\\ %
\scriptsize The Chinese University of Hong Kong
\and Benjamin Tan Kuan Wei\thanks{e-mail:working.celery@gmail.com }\\ %
\scriptsize National University of Singapore
\and Qingyang Xu\thanks{e-mail:qingyangxu@uchicago.edu }\\ %
\scriptsize The University of Chicago
\and Haiming Liu\thanks{e-mail:h.liu@soton.ac.uk }\\ %
\scriptsize University of Southampton
\and Shengdong Zhao\thanks{e-mail:shengdong.zhao@cityu.edu.hk}\\ %
\scriptsize City University of Hong Kong
 }

\teaser{
  \centering
  \includegraphics[width=\linewidth]{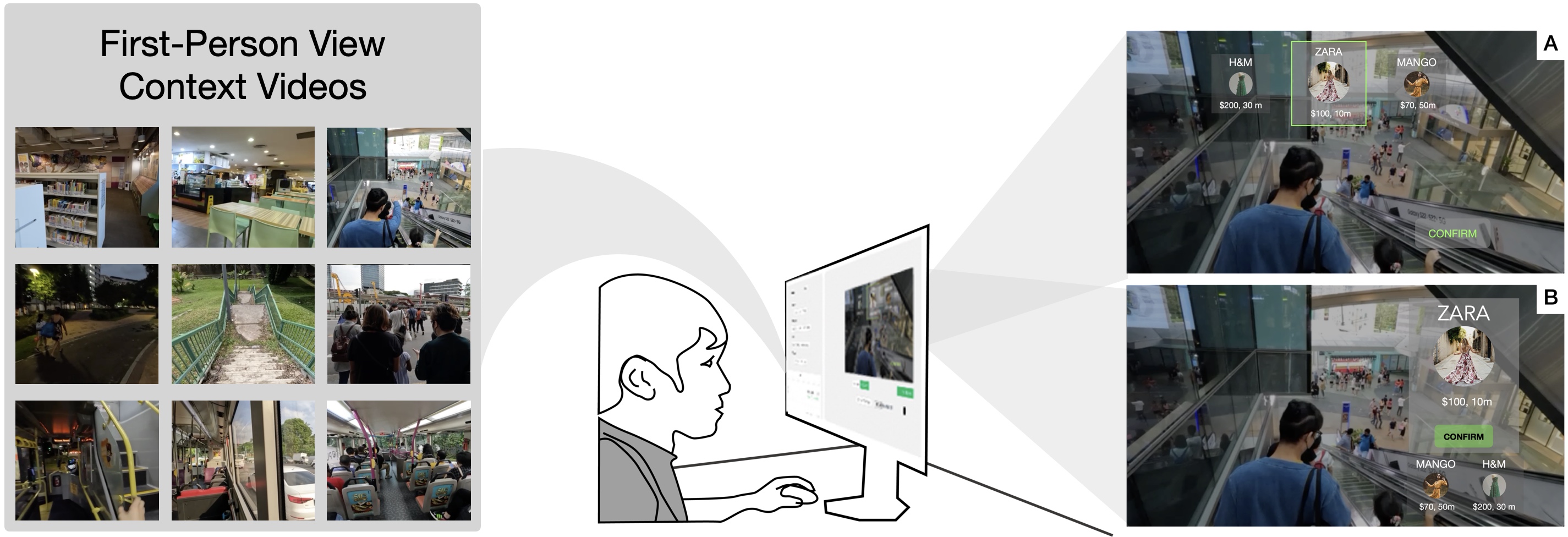}
  \caption{Depicts a scenario where designers want to test various potential designs for assisted reality (aR) based shopping UI on smart glasses. Our \vid{} approach supports rapid aR prototyping by combining first-person-view context videos with design prototypes to create design-blended videos that simulate the viewing experience of smart glasses.}
  \label{fig:teaser}
}

\abstract{
    Assisted Reality (aR) is a subfield of Augmented Reality (AR) that overlays information onto a user's immediate view via see-through head-mounted displays (OST-HMDs). This technology has proven to be effective and energy-efficient to support the user and information interaction for everyday wearable intelligent systems. The aR viewing experience, however, is affected by varying real-world backgrounds, lighting, and user movements, which makes designing for aR challenging. Designers have to test their designs in-situ across multiple real-world settings, which can be time-consuming and labor-intensive. We propose SimulataR, a cost-effective desktop-based approach for rapid aR prototyping using first-person-view context videos blended with design prototypes to simulate an aR expereince. A field study involving 12 AR users comparing SimulataR to real OST-HMDs found that SimulataR can approximate the aR experience, particularly for indoors and in low-to-moderate lit outdoor environments. Case studies with two designers who used SimulataR in their design process demonstrates the potential of design-blended videos for rapid aR prototyping.
} 


\newcommand{\vid}[0]{Simulat\emph{aR}}
\newcommand{\glass}[0]{OST-HMD viewing}

\begin{document}


\firstsection{Introduction}

\maketitle

\label{intro}

Assisted Reality (aR) is a type of Augmented Reality (AR) where information is overlaid in user's immediate view via see-through head-mounted displays (OST-HMDs) \cite{rauschnabel2022xr}. Unlike mixed reality (the more well-known form of AR) which integrates virtual objects seamlessly into the real world, aR projects a 2D virtual layer at a fixed depth. Although simpler, aR has demonstrated potential as an energy-efficient display method for wearable intelligent systems \cite{shen2023headsup}, gaining traction in both research \cite{chua2016position, ghosh2020eye, klose2019text, lu2020glance, rzayev2020position, ram2021lsvp, janaka2022para} and industry \cite{ptc2021, ama2022, realwear2022, vuzix2022, nreal2022, huawei2022}.

Given aR’s infancy, rapid prototyping tools are crucial to accelerate design evolution \cite{davies2005guest, beaudouin2007prototyping}. However, standard desktop-based tools like Figma or PowerPoint aren't tailored for rapid aR prototyping as these tools do not capture how content appears on a see-through OST-HMD display. More specifically, the visual appearance of the virtual content is affected by dynamic real-world backgrounds, lighting, and even the OST-HMD technology \cite{kruijff2010perceptual}. Consequently, designers must repeatedly transfer their prototypes to OST-HMDs for in-situ evaluation, a process that slows design cycles and raises costs. Furthermore, some scenarios, such as nighttime or public transport, may not be readily available for in-situ testing.


What if we could simulate a realistic aR experience for quick prototype assessment? Current approaches are limited. While some use CAVE systems \cite{gabbard2006text, ens2014cockpit} or VR headsets \cite{lee2013realism} to simulate mixed reality, they focus on simulating the stereoscopic aspect of an AR experience rather than how the virtual content appears in real-world settings, making them costly and complex to use.

In this work, we introduce \vid{}, a desktop-based rapid aR prototyping approach using design-blended videos. The design-blended videos are created by overlaying aR designs onto first-person view (FPV) context videos captured at eye-level. Adjustments, like transparency and field of view corrections, are made to design and context videos to mimic the OST-HMD perspective. Designers can assess and iterate designs for various scenarios by blending them with appropriate context videos.

To validate the feasibility of the proposed approach, we conducted a within-subject field study with 12 AR users, collecting over 108 hours of data. Participants rated the suitability of designs using OST-HMDs and \vid{} in diverse real-world settings. We covered various factors like location, lighting, mobility, and environmental artifacts. We also tested the scalability of \vid{} for simulating different OST-HMDs experience (Nreal, Hololens 2) and design types (simple notifications, and complex interfaces).

Results show that \vid{} effectively simulates aR in indoor and low-lit outdoor environments. However, in high-light situations (such as outdoor areas on a bright sunny day), especially with Hololens 2, effectiveness decreased. We further evaluated the potential of the approach for rapid prototyping through case studies with two designers. Qualitative findings suggest that \vid{} can be valuable during the initial stages of design for quickly narrowing down on UI design properties for real-world testing.

The contribution of this paper include:
\begin{itemize}
    \item An empirical field study to understand how well design-blended videos viewed on a desktop can simulate the viewing experience of real OST-HMDs.
    \item Preliminary insights from case studies on the potential of design-blending as a rapid prototyping approach for assisted reality interface design and validation.
\end{itemize}

\section{RELATED WORK}
\label{related}

\begin{table*}[h]
\caption{Comparison of AR simulation approaches}
\label{tab:related}
\resizebox{\linewidth}{!}{

\begin{tabular}{llllccc}
\hline
\textbf{\begin{tabular}[c]{@{}l@{}}Simulation\\ technique\end{tabular}} &
  \textbf{Target Experience} &
  \textbf{Purpose} &
  \textbf{\begin{tabular}[c]{@{}l@{}}Simulation\\ mechanism\end{tabular}} &
  \textbf{\begin{tabular}[c]{@{}c@{}}Low-cost\\ apparatus\end{tabular}} &
  \textbf{\begin{tabular}[c]{@{}c@{}}Supports non-experts\\ lacking coding skills\end{tabular}} &
  \textbf{\begin{tabular}[c]{@{}c@{}}Support simulation in \\ diverse real-world settings\end{tabular}} \\ \hline
{\color[HTML]{000000} De Sa et al. \cite{desa2012mobilear}} &
  Mobile AR &
  \begin{tabular}[c]{@{}l@{}}Rapid prototyping\end{tabular} &
  \begin{tabular}[c]{@{}l@{}}Video Prototype\\ on phone\end{tabular} &
  \ding{51} &
  \ding{51} &
  {\color[HTML]{000000} \ding{55}} \\ \hline
    Google AR Core \cite{googlear2021} &
  Mobile AR &
  \begin{tabular}[c]{@{}l@{}}Rapid prototyping\end{tabular} &
  \begin{tabular}[c]{@{}l@{}}Video Prototype\\ on phone\end{tabular} &
  \ding{51} &
  \ding{51} &
  \ding{55} \\ \hline
  {\color[HTML]{000000} Guven et al. \cite{guven2003author3d}} &
  Mixed Reality &
  \begin{tabular}[c]{@{}l@{}}Simulation environment\\ for controlled studies\end{tabular} &
  \begin{tabular}[c]{@{}l@{}}CAVE w/ \\ OST-HMD\end{tabular} &
  \ding{55} &
  \ding{55} &
  {\color[HTML]{000000} \ding{55}} \\ \hline
  Gabbard et al. \cite{gabbard2006text} &
  Mixed Reality &
  \begin{tabular}[c]{@{}l@{}}Simulation environment\\ for controlled studies\end{tabular} &
  \begin{tabular}[c]{@{}l@{}}CAVE w/ \\ OST-HMD\end{tabular} &
  \ding{55} &
  \ding{55} &
  \ding{55} \\ \hline
{\color[HTML]{000000} \begin{tabular}[c]{@{}l@{}} Lee et al. \cite{lee2013realism}\cite{lee2010latency}\end{tabular}} &
  Mixed Reality &
  \begin{tabular}[c]{@{}l@{}}Simulation environment\\ for controlled studies\end{tabular} &
  VR Headset &
  \ding{55} &
  \ding{55} &
  {\color[HTML]{000000} \ding{55}} \\ \hline

  \begin{tabular}[c]{@{}l@{}}\color[HTML]{000000} Ens et al. \cite{ens2014cockpit}\\ Ragan et al. \cite{ragan2009sim} \end{tabular} &
  Mixed Reality &
  \begin{tabular}[c]{@{}l@{}}Simulation environment\\ for controlled studies\end{tabular} &
  \begin{tabular}[c]{@{}l@{}}CAVE w/\\ head tracking\end{tabular} &
  \ding{55} &
  \ding{55} &
  {\color[HTML]{000000} \ding{55}} \\ \hline
\textbf{\vid{}} (Ours) &
  Assisted Reality &
  \begin{tabular}[c]{@{}l@{}}Rapid prototyping \\ for OST-HMDs\end{tabular} &
  \begin{tabular}[c]{@{}l@{}}View design-blended\\ videos on a desktop\end{tabular} &
  \ding{51} &
  \ding{51} &
  \ding{51} \\ \hline
\end{tabular}

}
\end{table*}

Our work examines the potential of using first-person-view (FPV) context videos for blending designs to simulate assisted reality, fostering rapid prototyping. This research leverages previous work in Rapid AR prototyping, AR simulation for OST-HMDs, and first-person-view video datasets.

\subsection{Rapid Prototyping for AR}

Rapid prototyping is crucial in the early design process, helping test ideas at low cost before finalizing interactive systems \cite{beaudouin2002prototyping}. Following a recently proposed XR framework \cite{rauschnabel2022xr}, we categorize the various AR prototyping approaches \cite{freitas2020systematic} based on the type of AR experience they can prototype.

\subsubsection{Rapid Prototyping for Mixed Reality} 
\emph{Mixed reality} experiences seamlessly integrate digital objects into our real-world view using tracking and mapping. Most existing AR prototyping approaches provide support towards this end of the AR space.

\textbf{Sketch-based prototyping:} These approaches enable users to embed digital sketches or paper prototypes into the user's view for mixed reality prototyping. DART \cite{macintyre2004dart}, an early tool, allowed users to create and embed sketches or 3D objects into the user's view using an OST-HMD but required coding skills. Recent works \cite{park2011arroom, gasques2019pintar, finke2019prototype} simplify this process by enabling users to import digital sketches directly into their augmented view for interaction. Nebeling et al. expanded paper prototyping to mixed reality, overlaying paper sketches and quasi-3D content from clay models to create mobile mixed reality experiences \cite{nebeling2018proto, nebeling2019proto}. 

\textbf{Video prototyping:} This is a low-cost method to convey interactive experiences swiftly by presenting designs in actual usage contexts for rapid feedback \cite{mackay1988video, mackay1999video}. Prototyping systems like Pronto permit spatially-aware placement of 3D sketches on 2D videos, enhancing mobile AR experiences \cite{leiva2020pronto}. Meanwhile, tools like Montage and Rapido foster design repurposing and transformation of non-interactive prototypes into immersive AR experiences \cite{leiva2018montage, leiva2021rapido}. Commercially, Google AR Core \cite{googlear2021} aids developers in recording and embedding 3D assets in real-world scenes.

In summary, while rapid mixed reality prototyping tools prioritize illustrating spatial and interactive experiences for designers, our focus is on assisted reality to offer designers a realistic preview of how their designs will appear on a real OST-HMD across real-world settings.

\subsubsection{Rapid Prototyping for Assisted Reality}

Assisted reality overlays digital content as a 2D virtual layer at a fixed depth in users’ real-world view. Although simpler than mixed reality, assisted reality has demonstrated potential as an energy-efficient display method to seamlessly support everyday computing tasks  and has gained traction among both researchers \cite{han2013asr, mura2016asr, klose2019text, lu2020glance, rzayev2020position, ram2021lsvp, sapkota2021ubi, ram2022dynamic, shen2023headsup, tan2023mindful, tan2024audiox, zhou2024glass} and industry \cite{ptc2021, ama2022, realwear2022, vuzix2022, nreal2022, huawei2022}.

Despite its importance, approaches for prototyping assisted reality are limited. 3D-HUDD prototypes 3D UIs for Heads-Up Displays (HUDs) on automobiles \cite{broy2015hudd} using static backgrounds. De Sa et al.  \cite{desa2012mobilear} simulate mobile AR by overlaying digital content on pre-recorded context video. However, assisted reality on OST-HMDs differs from mobile AR with the visual appearance of content being dependent on the context \cite{kruijff2010perceptual}.

In this work we explore the feasibility of blending designs onto the context video to simulate assisted reality experiences for OST-HMDs.

\subsection{Simulating AR experiences for OST-HMDs}
As outlined in Table ~\ref{tab:related}, previous work has utilized intricate setups such as CAVE-like domes or VR headsets to simulate mixed reality for OST-HMDs and often necessitate specialized equipment. For instance, some approaches require users to wear either an OST-HMD or stereo-tracking glasses and stand inside a hemispherical dome which is then projected with recordings of outdoor scenes to simulate mixed reality in specific settings \cite{guven2003author3d, gabbard2006text, ens2014cockpit, ragan2009sim}. While such intricate setups are warranted for supporting experts in conducting controlled experiments in a reproducible manner, we diverge by proposing a low-cost solution tailored for non-expert developers. Our approach involves utilizing design-blended videos to approximate assisted reality, enabling swifter evaluations and streamlining real-world testing processes.

\section{Study: Exploring the Feasibility of Simulating Assisted Reality on a Desktop with Design-Blended videos}
\label{study}

This study aims to determine if blending designs onto FPV context videos and viewing them on a desktop monitor, can realistically simulate assisted reality experiences. If so, it could expedite design evaluations for designers and enable remote assessments across different contexts, particularly benefiting those without easy access to OST-HMD platforms. 

Our primary research question is: To what extent can design-blended videos simulate the OST-HMD-based assisted reality for different UIs, contexts, and OST-HMD platforms ?


\subsection{Method}

To answer our research question, we designed a field study where AR designers compared and rated UI designs using two viewing methods: 1) a real OST-HMD in a specific setting (\glass{}), and 2) a design-blended FPV video of the identical setting (with corrections as described in Sec~ \ref{video_prototype}) on a desktop monitor (\vid{}). Consistent ratings across both methods would indicate that \vid{} effectively simulates the OST-HMD viewing experience in that setting.

\begin{figure*}[t]%
\centering
\includegraphics[width=\linewidth]{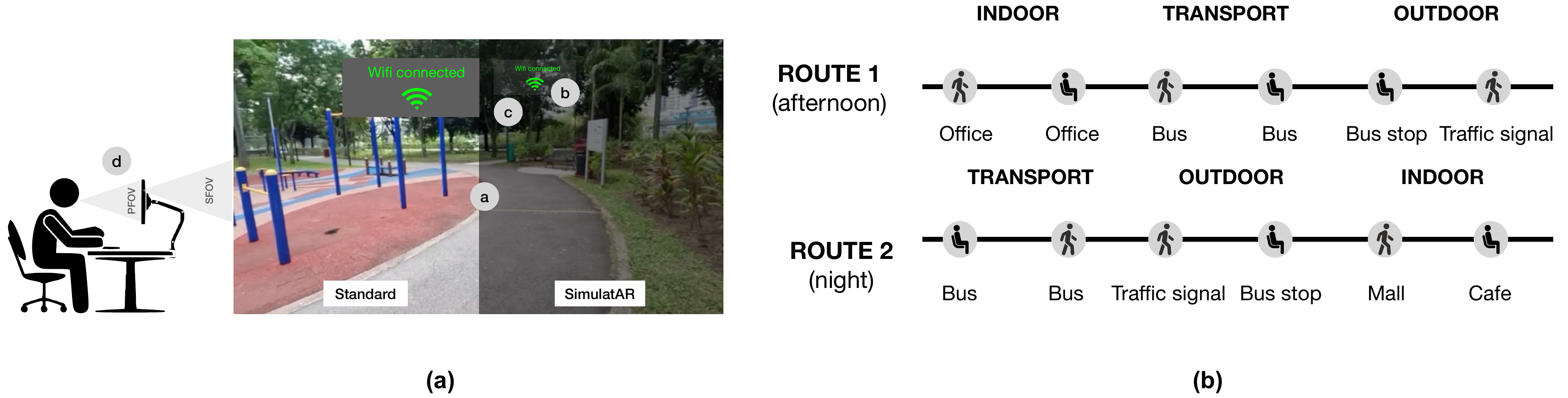}
\caption[]{(a) Comparison of a simple overlay of content on a context video (left half) and a design-blended video with visual corrections (right half) and (b) An overview of the routes undertaken by the user}
\label{fig:proto}
\end{figure*}

\textbf{Context generalizability.}  Assisted reality is affected by contextual factors such as location, lighting level, environmental artefacts, and users' mobility. To test SimulataR's accuracy across these factors, we explore two routes (see Figure~ \ref{fig:proto}b) differing in lighting (low: $\approx$100 lux outdoor, $\approx$250 lux indoor; high: $\approx$10,000 lux outdoor, $\approx$500 lux indoor) \cite{eng2004ill}. Each route includes three different location settings (indoor, outdoor, and transport) with two common mobility tasks (sitting, walking).

\textbf{Platform generalizability.} A wearer’s OST-HMD-viewing experience is also dependent on the OST-HMD itself, with properties such as resolution, FOV, and display technology affecting users’ perception of content \cite{argelaguet2020review}. We thus consider two OST-HMD platforms (Nreal Light and Hololens 2) of different properties and compare them with their respective \vid{} alternatives.

\textbf{Content generalizability.} We consider how \vid{} can be used to evaluate text and graphics-based digital content typically encountered in mobile computing. We look into two types of designs of differing complexity: simple icon-based notifications and more complex text snippets with photorealistic images (e.g. news reading or social media interfaces).











\begin{table}[t]

\caption{Overview of OST-HMD platforms}
\resizebox{\columnwidth}{!}{
\begin{tabular}{@{}lll@{}}

\toprule
\textbf{}                     & \textbf{Microsoft Hololens 2 \cite{hl2022}} & \textbf{Nreal Light \cite{light2022}} \\ \midrule
\textbf{Resolution (per eye)} & 1440 x 936           & 1920 x 1080                   \\
\textbf{Optics}               & Waveguides           & Birdbath                      \\
\textbf{Display}              & Laser Beam Scanning  & OLED                          \\
\textbf{Support for assisted reality} & Dedicated mirroring mode & Requires custom application \\ \bottomrule

\end{tabular}
}

\label{tab:platforms}
\end{table}





\subsection{Design-Blended Videos}
\label{video_prototype}


To construct design-blended videos that closely resemble real OST-HMD experiences, we conducted informal pilots to identify key corrections needed for videos with a simple design overlay. In the pilots, participants assessed sample designs consisting of text and solid backgrounds on Nreal Light and Hololens 2 platforms at common low (100 lux outdoor, 250 lux indoor) and high (10000 lux outdoor, 500 lux indoor) lighting levels \cite{eng2004ill}. After viewing, they adjusted the visual properties of the design to match OST-HMD experience more accurately. Two primary corrections were pinpointed:

\begin{enumerate}
    \item Additive light corrections: To simulate the additive nature of OST-HMD displays, we first overlay a black tinted window of appropriate opacity on the FPV videos to match the tint level of the OST-HMD (Figure~ \ref{fig:proto}a). We then adjust the opacity level of any solid backgrounds in the design based on the external lighting levels of the simulated context (Figure~ \ref{fig:proto}b).

    \item FOV corrections: To align with the FOV of the OST-HMD and enhance immersion (Figure ~\ref{fig:proto}c, ~\ref{fig:proto}d), we adjusted the scale of the overlaid content and seated users such that their physical FOV with the desktop monitor matched the software FOV captured by a GoPro camera (1:1 ratio), following previous research  \cite{draper2001fov, mary2002fov, poly2005fov}.

\end{enumerate}

These corrections are applied to the design and context videos before superimposing them to create the design-blended videos for our study.

\subsection{Experiment Design}

The study is within-subject with participants evaluating designs using the two viewing methods (OST-HMD viewing, \vid{}) blocked by the OST-HMD platform (Nreal, HL2) for two routes. For each location-mobility pair in a route, participants evaluated two types of content (simple, and complex). Following Oulasvirta et al \cite{oulasvirta2005rcf}, we tackled order effects by randomizing the route direction (normal or reverse), order of OST-HMD platform, and order of content type. The order of viewing methods was fully counterbalanced. 

\subsubsection{Stimuli}

We recorded context videos in real-world locations that are representative of the contexts shown in Figure ~\ref{fig:proto}b. Our recording setup is similar to the one in \cite{huang2022immersive} and utilized a GoPro Hero10 camera affixed 5 cm in front of the users' eyes on a bicycle helmet. The camera operated at 50 FPS with a 2704 x 1520 (2K) resolution and a 95° diagonal FOV in Linear mode (16:9 aspect ratio) \cite{gopro2021specs}, leveraging standard Hypersmooth stabilization to simulate the vestibulo-ocular reflex stabilization effect \cite{fetter2007vestibulo}.

For each design, we created two design variants conveying the same information but differing in visual properties that can impact usability on OST-HMDs such as background, text properties such as size, font, and color \cite{gabbard2006text, ram2021lsvp}, and position and layout in users’ FOV \cite{rzayev2018reading, ghosh2020eye, janaka2022para}. The designs were superimposed on context videos by making the necessary corrections as described in Sec. \ref{video_prototype}.

\subsection{Measures}

To understand the similarity of the viewing methods, users are asked to subjectively assess each design variant in terms of its suitability for use on OST-HMDs in a given context. Users assessed various design dimensions used in prior OST-HMD literature namely noticeability, identifiability, comfort, level of environmental awareness, and multitaskability \cite{janaka2022para, chua2016position, jankowski2010text} using a 7-point Likert scale. We compare user ratings for each design variant in either viewing methods to understand the similarity in user’s perception between \vid{} and OST-HMD. High similarity (statistically equivalent) indicates that the viewing methods are equivalent. In addition, users' subjectively compare the overall similarity between \vid{} and \glass{} for each context.

\subsection{Study Apparatus}

For OST-HMD-viewing we consider two OST-HMD platforms, Nreal Light (Nreal) and Hololens 2 (HL2), that differ in their specifications. For the \vid{} condition, the design-blended videos were displayed on a desktop monitor centered at the user's eye level. Users were seated 30 cm from the display to match the physical and software FOV. 

\subsection{Participants}

We recruited 12 users (7 Female, 5 Male; M=25.54, SD=2.63 years) who had experience with OST-HMD platforms. Six had 3-5 years of experience with building developing AR apps for smartphones and OST-HMDs using Unity. The other six had 1-2 years of OST-HMD content design or had participated in multiple research studies using OST-HMDs in the past. 

Participants evaluated design variants across 12 scenarios (6 per route) over two days, taking on average 9 hours to complete the study. On a given day, participants completed both viewing conditions for an OST-HMD in a two hour session with a break of 30 minutes between the conditions. They received $60$ USD for their participation. This study was approved by the IRB of our institution, and an informed consent was obtained from every participant.

\subsection{Study Procedure}

The study started with task explanation and a short training for accessing the designs on the OST-HMD platform for that day. They then sequentially evaluated design variants for each content type in the assigned viewing method as they progress through the route, providing immediate subjective ratings and choosing a preferred variant. After each session, a short semi-structured interview was held to gather insights on the pros and cons of each viewing method.

\section{Study: Results}
\label{results}

\subsection{Data Analysis}

We gathered 5760 data points over 108 experimental hours. Due to violations of ANOVA's normality assumption, we used aligned rank transformation \cite{wobbrock2011art} followed by Tukey corrected post hoc tests for analysis. When no significant differences were noted, we employed the Two One-Sided Test (TOST) \cite{schuirmann1987tost} to check for viewing method equivalence, defining a $\pm 1$ difference in the rating scale range as practically equivalent to zero.

\begin{figure*}[t]%
\centering
\caption[]{Descriptive table of equivalence in users’ assessments for the design dimensions in each context. Each cell is marked green if there is significant equivalence of ratings for both design variants, yellow if equivalence was seen only for one of the variants, and red if equivalence failed for both variants.}

\includegraphics[width=\textwidth]{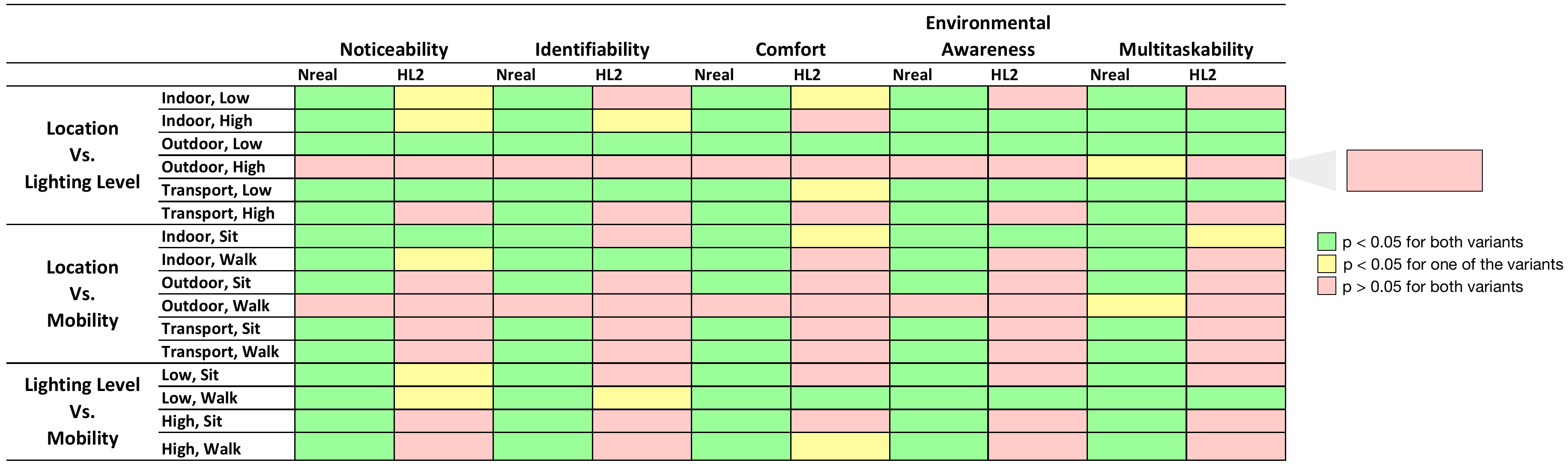}

\label{fig:equivalence}
\end{figure*}

We present the results in the format of design variant A stats followed by variant B stats (i.e., $(p=0.1)_{A}; (p=0.5)_{B}$ indicates a $p$ value of $0.1$ for A variant, and $p$ value of $0.5$ for B variant). See Figure ~\ref{fig:equivalence} for a summary of how context affects \vid{} and OST-HMD viewing equivalency.

\subsection{Similarity of assessments between viewing methods}

\subsubsection{Nreal} A repeated measures ANOVA did not show any significant differences between viewing methods for either design variant and were statistically equivalent for noticeability $(F_{1,517}=0.25, p=0.61)_{A}(t(11)=5.93, p<0.001)_{A}; (F_{1,517}=0.03, p=0.86)_{B}(t(11)=5.59, p<0.001)_{B} $, identifiability $(F_{1,517}=0.04, p=0.83)_{A}(t(11)=5.33, p<0.001)_{A}; (F_{1,517}=0.31, p=0.57)_{B}(t(11)=5.79, p<0.001)_{B} $, comfort $(F_{1,517}=0.07, p=0.78)_{A}(t(11)=6.7, p<0.001)_{A}; (F_{1,517} = 0.19 , p=0.65)_{B} (t(11)=5.78, p<0.001)_{B}$, environmental awareness $(F_{1,517}=0.04, p=0.84)_{A}(t(11)=5.87, p<0.001)_{A}; (F_{1,517}=0.03, p=0.85)_{B}(t(11)=5.57, p<0.001)_{B} $, and multitaskability $(F_{1,517}=0.22, p=0.63)_{A}(t(11)=8.03, p<0.001)_{A}; (F_{1,517}=0.09, p=0.76)_{B}(t(11)=6.07, p<0.001)_{B} $.

The ANOVA revealed no significant interactions for viewing method x mobility, and viewing method x design type across all design dimensions. However, significant interactions were found for viewing method x lighting level for noticeability $(F_{1,517}=4.46, p=0.03)_{B}$, and for viewing method x location for noticeability  $(F_{1,517}=4.09, p=0.01)_{B} $ and comfort $(F_{1,517}=5.13, p<0.001)_{B} $ for one of the design variant. Despite this, post-hoc tests showed no significant differences, except in highly lit outdoor locations and during outdoor walking scenarios.

\subsubsection{Hololens 2} A repeated measures ANOVA showed significant differences between viewing methods for noticeability $(F_{1,517}=78.1, p<0.001)_{A}; (F_{1,517}=51.06, p<0.001)_{B} $, identifiability $(F_{1,517}=60.84, p<0.001)_{A}; (F_{1,517}=73.90, p<0.001)_{B}$, comfort $(F_{1,517}=40.49, p<0.001)_{A}; (F_{1,517}=34.83, p=<0.001)_{B}$, environmental awareness $(F_{1,517}=18.92, p<0.001)_{A}; (F_{1,517}=4.03, p=0.04)_{B}$, and multitaskability $(F_{1,517}=15.54, p<0.001)_{A}; (F_{1,517}=4.05, p=0.04)_{B}$, suggesting that \vid{} did not simulate a HL2 viewing experience well enough to assess the designs. 

Significant interactions were also found for all combinations of viewing method with other contextual factors for the design dimensions. Post hoc contrasts and equivalence tests showed that assessments were non-equivalent primarily at high lighting levels across location, mobility, and content type.

\begin{table}[t]
\caption{Descriptive results for the similarity of \vid{} to \glass{} based on contextual factors. The average similarity values (1=Dissimilar, 7=Similar) are color coded; with green indicating high levels of similarity, yellow indicating sufficient similarity, and red indicating a low similarity to \glass{}. }
\resizebox{\columnwidth}{!}{
\begin{tabular}{clrrrrrr}
\hline
\multicolumn{2}{l}{} &
  \multicolumn{2}{c}{\textbf{Indoor}} &
  \multicolumn{2}{c}{\textbf{Outdoor}} &
  \multicolumn{2}{c}{\textbf{Transport}} \\ \hline
\multicolumn{2}{l}{} &
  \multicolumn{1}{c}{\textbf{Sitting}} &
  \multicolumn{1}{c}{\textbf{Walking}} &
  \multicolumn{1}{c}{\textbf{Sitting}} &
  \multicolumn{1}{c}{\textbf{Walking}} &
  \multicolumn{1}{c}{\textbf{Sitting}} &
  \multicolumn{1}{c}{\textbf{Walking}} \\ \hline
\multicolumn{1}{c|}{} &
  \textbf{Nreal} &
  \cellcolor[HTML]{7BFF7C}5.92 (0.90) &
  \cellcolor[HTML]{7BFF7C}5.67 (0.98) &
  \cellcolor[HTML]{FFFE65}4.92 (1.16) &
  \cellcolor[HTML]{FFCCC9}3.92 (1.44) &
  \cellcolor[HTML]{7BFF7C}5.42 (1.00) &
  \cellcolor[HTML]{FFFE65}4.33 (1.23) \\ \cline{2-2}
\multicolumn{1}{c|}{\multirow{-2}{*}{\textbf{\begin{tabular}[c]{@{}c@{}}High \\ Lighting\end{tabular}}}} &
  \textbf{HL2} &
  \cellcolor[HTML]{7BFF7C}5.00 (1.13) &
  \cellcolor[HTML]{FFFE65}4.42 (1.08) &
  \cellcolor[HTML]{FFCCC9}3.58 (1.31) &
  \cellcolor[HTML]{FFCCC9}2.58 (0.79) &
  \cellcolor[HTML]{FFCCC9}3.50 (1.45) &
  \cellcolor[HTML]{FFCCC9}2.58 (1.00) \\ \hline
\multicolumn{1}{c|}{} &
  \textbf{Nreal} &
  \cellcolor[HTML]{7BFF7C}6.17 (0.94) &
  \cellcolor[HTML]{7BFF7C}5.67 (0.98) &
  \cellcolor[HTML]{7BFF7C}6.00 (0.85) &
  \cellcolor[HTML]{7BFF7C}5.17 (1.11) &
  \cellcolor[HTML]{7BFF7C}5.42 (1.68) &
  \cellcolor[HTML]{FFFE65}4.75 (1.48) \\ \cline{2-2}
\multicolumn{1}{c|}{\multirow{-2}{*}{\textbf{\begin{tabular}[c]{@{}c@{}}Low \\ Lighting\end{tabular}}}} &
  \textbf{HL2} &
  \cellcolor[HTML]{7BFF7C}5.25 (1.29) &
  \cellcolor[HTML]{FFFE65}4.83 (0.94) &
  \cellcolor[HTML]{7BFF7C}5.08 (1.16) &
  \cellcolor[HTML]{FFCCC9}3.92 (1.31) &
  \cellcolor[HTML]{FFFE65}4.50 (1.57) &
  \cellcolor[HTML]{FFCCC9}3.67 (1.50) \\ \hline
\end{tabular}
}
\label{tab:sim_scores}
\end{table}

\subsection{Similarity of \vid{} Vs. \glass{}}

Table \ref{tab:sim_scores} shows average ratings on how similar \vid{} was to \glass{} in different contexts for each platform. In general, we see that \vid{} closely approximates \glass{} at low lighting levels. Notably, it offered greater stability with Nreal even in high lighting compared to HL2. However, at high lighting levels, \vid{}'s usability declined, particularly on the HL2 platform, with users' mobility further impacting the simulation experience.

\subsection{Qualitative Feedback} 

There was general consensus that \vid{} can be an efficient and useful technique for rapid prototyping of assisted reality. Users suggested that \vid{} could be particularly useful in the early stages of design, helping narrow down on potential designs for extensive real-world testing: "It [\vid{}] can save me a lot of time... I can cut down 70-80\% of the bad designs and only check the really good ones on the glass." (P3). Furthermore, users also suggested that \vid{} can be potentially useful for newcomers in the AR/MR design field: "This would have helped me a lot when I started out... I can quickly learn how designing for OST-HMDs is different, what works and what doesn't" (P6).

Moreover, \vid{} helped address the issue of social discomfort experienced during public testing with wearables \cite{denning2014priv, rico2009priv}, as cited by two participants, allowing for a more focused and comfortable evaluation process without the scrutiny of bystanders. For instance, P4 contrasted their earlier experience of evaluating with an OST-HMD in public with \vid{}, "[With real HL2] Everyone was looking at me when I was wearing the Hololens and making gestures. It's very uncomfortable for me and maybe for them also... I try to finish my testing as fast as possible... It affects my focus on testing... [using \vid{}] I can think about the issues in my design more clearly" (P4). Thus, \vid{} facilitates efficient, convenient, and discrete design assessments from a desktop.

\subsection{Discussion}

Our study explored the effectiveness of using SimulataR to assess UIs designed for assisted reality experiences in various settings. We now reflect back on the research question we posed earlier and discuss our findings. Overall, we see that \vid{} can successfully simulate  the \glass{} experience, albeit with varying degrees of accuracy depending on contextual factors, with lighting level having the highest impact followed by the mobility task.


From Table \ref{tab:sim_scores} we see that at low to moderate lighting levels, \vid{} provided a successful simulation of the \glass{} experience across contextual factors, OST-HMDs, and design types, supporting our case for the generalisability of \vid{}. This finding is validated by the statistical equivalence of design ratings and users' comments: "The transparency and the layout of the content are captured really well on the simulator" (P2), [Comparing \glass{} to \vid{}] The way the content shakes and my attention split is not exactly the same,... but still the simulation gives me an overall idea of how the experience will be" (P3). 


However, \vid{} was less successful for design assessment in high lighting conditions, especially in outdoor and transport settings, as indicated by the significant difference in ratings in these situations. This could be due to two reasons. First, in bright outdoor settings, virtual imagery on the OST-HMD appears “washed-out” with reduced contrast \cite{argelaguet2020review, erickson2020light}. Although we corrected for this effect during design blending, users felt that \vid{} overestimated the visibility compared to \glass{}. Second, the real-world high lighting induces changes in pupil size, which affects visual acuity and contrast sensitivity \cite{strang1999defocus}. Such physiological reactions are challenging to mirror using \vid{}. Moreover, these limitations were more pronounced when users were mobile, possibly because users could focus on and perceive virtual content even with reduced visibility when seated. 

Our results also indicate that at high lighting levels, \vid{} was better at assessing designs for NReal than HL2. One reason for this is that the HL2 platform has a poorer contrast ratio (2-3\%) than Nreal in bright ambient light \cite{erickson2020light}, which in turn makes the platform more prone to color blending effects \cite{kruijff2010perceptual}. In addition, HL2 has color uniformity problems compared to Nreal \cite{Karl2021nrealvshl2}, possibly due display technology variations. \vid{} did not adequately capture these contrast and color issues in HL2, signaling the necessity for further adjustments to \vid{} facilitate evaluations in these situations.

In summary, our results indicate that SimulataR can be used to assess assisted reality designs in a generalisable manner in low to moderate lighting levels across real-world settings, but it may not capture the true assisted reality experience in outdoor high lighting situations, especially for HL2.

\section{Case Studies with Designers}
\label{tool}

Our field study revealed that design-blended videos on a desktop have the potential to visually approximate wearable assisted reality experiences across diverse conditions. To understand whether such an approach could be useful to designers when designing wearable AR experiences, we instantiated the \vid{} approach as a web-based tool that creates design-blended videos using prototypes provided by the user. We used the tool as a probe to investigate the the rapid prototyping capabilities of the design-blending approach through case studies with two designers.

\subsection{\vid{} Tool}

The tool consists of two primary components: a set of context videos for use in the prototyping process and a web interface to facilitate design-blending.


\textbf{Context Video Pool} To allow designers to readily use our tool for design assessment, we decided to expand upon the pool of context videos collected in the field study. In collecting new FPV videos, our objective was to cover more representative contexts where wearable displays will be likely be used. Guided by a recent taxonomy describing common usage contexts of wearables \cite{akpinar2020small}, we captured diverse contexts (e.g., commute in bus, walking in a mall, sitting in a park) in different lighting conditions.



\begin{figure}[]%
\centering

\includegraphics[width=\columnwidth]{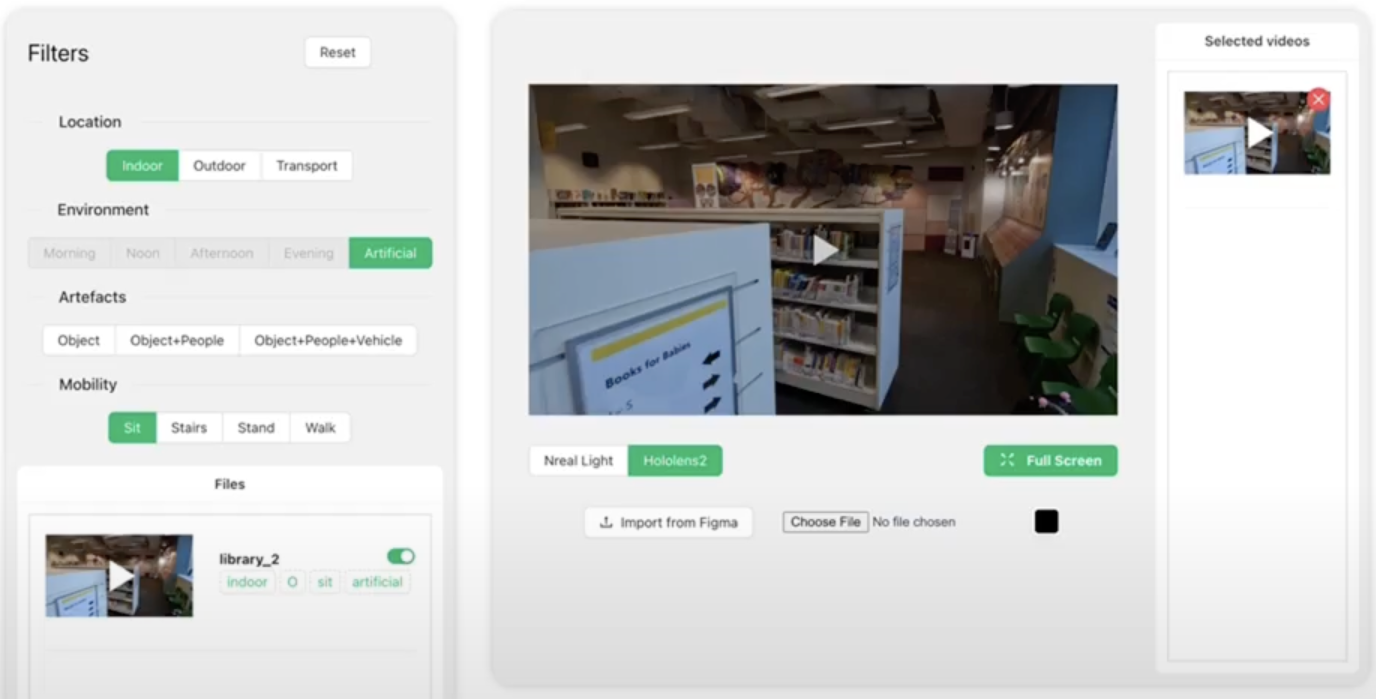}

\caption[]{Design-blending tool used for the case study}

\label{fig:tool}
\end{figure}

\textbf{Web Interface} We created a basic interface (refer to Figure \ref{fig:tool}) enabling designers to select a context video from our collection and specify the smart glasses they want to simulate (HL2 or Nreal). They can then upload their design prototype, created either in Figma or other external tools, as an image. Later, they can view the resulting design-blended video in full-screen mode. The blending of the design with the video is done automatically, adjusting the properties of both elements as detailed in Section \ref{video_prototype}.

\subsection{Participants}
We reached out to two designers who were interested in using the tool for their AR application design process. The first designer (D1) had 5 years experience designing for web and mobile interfaces. They had previously designed a mobile AR application, but not for wearable AR. The second designer (D2) had 3 years of web design experience and had recently designed a wearable AR application for composing and replying to emails across diverse mobile contexts (e.g. walking, bus/train commute).

\subsection{Findings}

\subsubsection{Tool Usage}


D1 employed the \vid{} tool to develop a mixed reality application intended to enhance students' learning from study materials, such as biology textbooks, by integrating augmented images and labels. They initially designed a prototype using the standard light gradient colors available in VR toolkit, but the prototypes proved problematic during tests on Hololens 2 due to low readability. To address this, D1 used \vid{} with a context video depicting an office scenario with a person seated at a desk to refine the UI. The UI improvements were carried out in two phases: the first adjustment focused on optimizing text readability and color, followed by modifications to the UI size and layout. The resulting design was found to be more effective on Hololens 2.


In D2's case, having already developed the wearable email app, they used \vid{} to retrospectively analyze how it could have influenced their design process. They tested their initial designs using \vid{} in diverse context videos since usability across different environments was crucial for their app. They identified two promising versions based on \vid{} output, which they felt had to be tested in real-world conditions before finalizing the design.

\subsubsection{Qualitative Feedback}

Both designers found \vid{} to be easy to use and valuable for early-stage prototyping. The design-blended videos were found to closely resemble the wearable viewing experience, offering a clearer understanding of how a particular context would affect the appearance of their UI prototypes. From participants' feedback, we identified three main use cases of \vid{}.

\emph{Color scheme and text design.} Both designers noted the tool's effectiveness in identifying UI color schemes and text properties that are clear and visually suitable for the intended usage context. The transparency and tint effects applied to the designs and context videos provided clear insights into what designs work and what doesn't. For instance, D2 contrasted their earlier design process for their AR email application, noting the significant reduction in time to identify appropriate text properties: "It [traditional AR design process] took a lot of time ... I tried multiple options as they all looked good on the desktop but were so different when I loaded in the smart glasses.... with \vid{} I could compare multiple text styles in the scene and I quickly found the same winning options as last time."

\emph{Spatial layout and size estimation.} Users reported that the context video was not only useful in adding a sense of realism to the design process, but also served as an reference to decide the spatial layout and size of UIs needed for their application. For instance, D2 found that the tool could be used to evaluate how different elements of the UI would block or interact with the user's view in real-world scenarios, such as in busy or constrained spaces like hallways, buses, and supermarkets.  D1 also identified another interesting application: they used it as a comparative tool to estimate UI sizes in Unity based on the size of various environmental objects in the context video, streamlining what would otherwise be an iterative task: "My context video has an office desk with some study notes in front of the user... [when using \vid{}] I realised that the [virtual] images in my application should roughly be the size of text block in the note.... I had a much better sense of each element's position and size".

\emph{Scalability to diverse contexts} SimulatAR proved particularly useful for designing applications intended for use in diverse settings. For instance, D2 highlighted how \vid{} facilitated remote evaluation of designs in multiple contexts, which could significantly reduce the effort and testing time: "with SimulatAR I could have tested my designs without physically travelling to each location... it [the application development] would have been much faster and easier".

\begin{figure*}[t]%
\centering

\includegraphics[width=\linewidth]{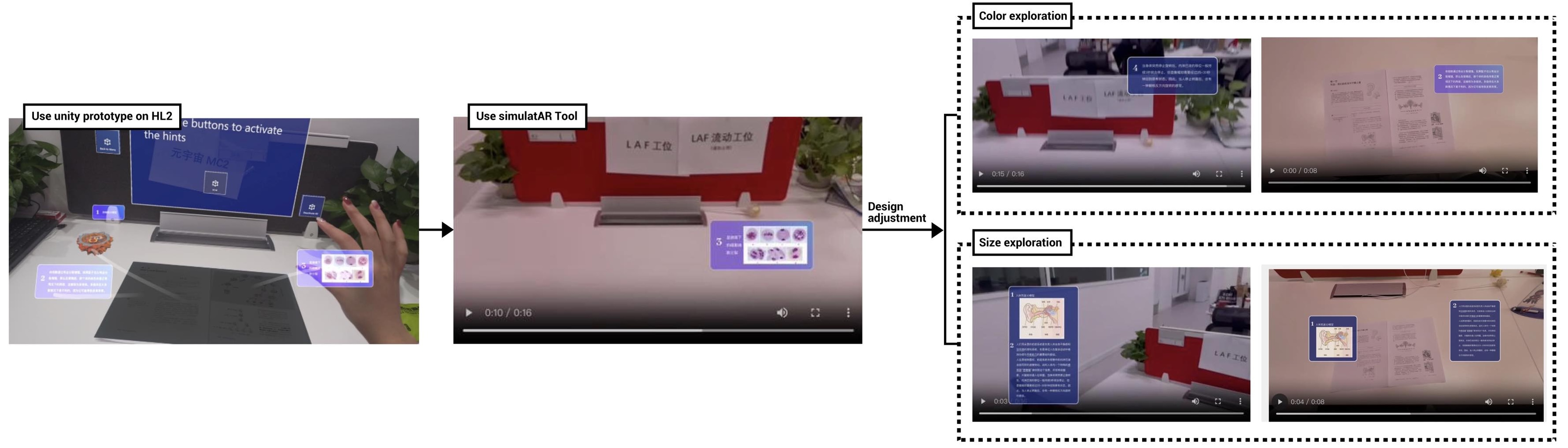}

\caption[]{The design process adopted by designer D1 in the case study. They began with testing on a Hololens 2, followed by testing on \vid{} to make adjustments in the color, size, and layout of the UIs. The designs were subsequently verified in diverse contexts.}

\label{fig:design_process}
\end{figure*}

\section{Discussion}

The field studies and case studies presented in this paper demonstrate the potential of design-blending as an early-stage rapid prototyping of assisted reality interfaces. Below, we discuss the key considerations that designers must look at when deciding to use \vid{} for their prototyping and how the research community can build on this approach to further enhance the prototyping experience.

\subsection{When should designers use \vid{} ?}

We envision \vid{} as a pivotal approach for early-stage rapid prototyping of assisted reality interfaces, enabling designers to evaluate AR application variants and shortlist designs for real-world testing. It is particularly effective for simulating assisted reality in indoor and moderately lit outdoor settings. However, its performance reduces in brighter settings, especially on the HL2 platform. Addressing this shortfall would involve refining the brightness and color correction models \cite{david2014color}. Moreover, integrating recent advances in generative modelling \cite{karras2020stylegan} could facilitate data-driven end-to-end solutions for creating design-blended videos conditioned on the \vid{} inputs.

Another use case of \vid{} that we foresee is its utilization as an educational tool to provide insights into the assisted reality experience. Many educational institutions lack access to AR glasses, limiting hands-on AR exposure. \vid{} can provide a preliminary exposure to AR environments, fostering innovative interactive designs beyond traditional mobile applications. It can offer students a better sense of AR interface design, preparing them for real-world AR glass scenarios, enhancing creativity, and advancing AR design education.

\subsection{When should we consider alternatives to \vid{} ?}

To simulate mixed reality prototypes involving spatial positioning or controlled studies where detailed metrics like response times need to be measured, high-fidelity techniques might be preferable. \vid{} compromises on depth perception and immersion due to its monoscopic approach, focusing instead on wearable assisted reality where depth simulation is less vital. This design choice facilitated affordable and simpler setups, even for users with minimal coding skills. Future research could investigate affordable alternatives to CAVE setups to offer users immersive experiences, potentially applying recent studies utilizing $360^{\circ}$ videos \cite{huang2022immersive} and spatially-aware multi-viewpoint videos \cite{leiva2020pronto} to enrich our design blending approach.

\subsection{Limitations}

The outdoor scenarios in our dataset and field study focus on tropical, humid regions. Though we propose using egocentric video datasets to diversify contexts, the effect of differing FOVs and video perspectives remains unclear. While we anticipate tolerance to moderate variations of these factors, future research should test the validity of our approach for these cases.

Additionally, our study's sample size $(n=12)$ is relatively small, as we specifically targeted AR users with prior OST-HMD experience. We chose this population as their experience is essential to judge the similarity between \vid{} and \glass{} and provide insightful feedback on even nuanced contrasts between the viewing methods. We anticipate that future research cab replicate and expand upon our findings

\section{Conclusion}
\label{conclusion}

We explored the feasibility of simulating the assisted reality offered by OST-HMDs on a desktop monitor using design-blended videos (\vid{}). Our evaluation with experienced AR users demonstrated that assessments made with \vid{} were statistically equivalent to assessments made using a real OST-HMD, highlighting its utility for quickly gauging assisted reality designs across a diverse range of contexts, OST-HMD platforms, and design types. Our findings contribute to the literature on rapid prototyping for AR, and we envision a pipeline to facilitate rapid prototyping and narrowing down for designs real-world testing.

\bibliographystyle{abbrv-doi}

\bibliography{main}

\end{document}